\documentclass[twocolumn,preprintnumbers,amsmath,amssymb]{revtex4}
\usepackage{epsfig}

\begin{document}

\title{Vacuum effects in a vibrating cavity: \\ time refraction, dynamical Casimir effect, and effective Unruh acceleration}

\author{J.T. Mendon\c{c}a}

\email{titomend@ist.utl.pt}

\affiliation{CFIF and GoLP, Instituto Superior T\'{e}cnico,1049-001 Lisboa, Portugal}

\author{G. Brodin and M. Marklund}

\affiliation{Department of Physics, Ume{\aa} University, SE--901 87 Ume{\aa}, Sweden}

\begin{abstract}
Two different quantum processes are considered in a perturbed vacuum cavity: time refraction and dynamical Casimir effect. They are shown to be physically equivalent, and are predicted to be unstable, leading to an exponential growth in the number of photons created in the cavity.  The concept of an effective Unruh acceleration for these processes is also introduced, in order to make a comparison in terms of radiation efficiency, with the Unruh radiation associated with an accelerated frame in unbounded vacuum.  

\end{abstract}

\maketitle

\section{Introduction}

Vacuum has always been an essential ingredient of our knowledge of the
physical world, from Aristotle to the present days. In the low energy limit
of the vacuum fluctuation spectrum, as described by quantum electrodynamics
(or qed), vacuum effects predict the emission of pairs of photons, induced
by some external perturbation. For higher energy fluctuations, qed also
predicts the occurrence of vacuum nonlinearities, which are associated with
virtual electron-positron pairs \cite{marklund-shukla}. At even higher energies, electron-positron pairs will
eventually be emitted from vacuum and become real. Other real and virtual
particle-antiparticle pairs also have to be considered.

Here we restrict our analysis to the low energy range of quantum
electrodynamics, where the influence of electron-positron pairs can be
neglected. This is the range of quantum optics, which only
deals with photon vacuum effects. It should be noted that the effects to be considered here
could also occur at higher energies, involving other particles and other
fields.

Three different effects have been discussed in the frame of photon qed or
quantum optics: i) dynamical Casimir effect \cite
{moore,dodonov1,dodonov2,law,uhlmann}, ii) Unruh-Davies radiation \cite
{unruh,davies}, and iii) time refraction \cite{mend1,mend2,mendpra}.
Dynamical Casimir effect is a direct extension of the famous double plate
geometry of the Casimir effect \cite{casimir}, which revealed the energy difference between different vacua. The dynamical Casimir setup considers one of the plates as periodically
oscillating in time, due to some applied force. Unruh-Davies radiation
(also called Unruh radiation) demonstrates the existence of thermal
radiation, as seen from an accelerated reference frame in unbounded vacuum.
At first sight this could be a purely kinematic effect, with no physical
consequences. Its real existence has actually been seriously questioned \cite
{ford}. However, any physical detector (such as a charged particle or an
atom) moving with the accelerated frame, would be able to interact with such
thermal radiation and thus respond accordingly. An important aspect of
the Unruh effect is that it explores the equivalence between gravitation and
acceleration, and is intimately related to Hawking radiation \cite
{hawking}. The relations between the dynamical Casimir effect and the Unruh
effect have been explored in, e.g., Refs. \cite{Eberlein1996,Milton2004}, and
various views as to how close the relationship are have been presented in
the literature \cite{Hu2004}. The interplay between these two vacuum effects have important
consequences for how experiments should be interpreted, see for example the
recent debate \cite{Scully2003,Hu-comment,Scully-reply}.

In the present paper we will attempt to shed further light on these
connections by tying the dynamical Casimir effect to the concept of time
refraction. Time refraction is the temporal version of the well known
concept of refraction. It is a low order effect, which is perceived by any
photon in a time varying medium, and can be seen as the most basic mechanism
leading to photon acceleration \cite{mendbook}. As a result of time
refraction, super-luminal frames with constant velocity can also observe a
radiation spectrum resembling the Unruh radiation \cite{ariel}.

In a recent work \cite{mendpra}, we were able to show that the concept of
time refraction, when considered in the specific case of an optical cavity,
is very similar to the dynamical Casimir effect. Time refraction always
involves the presence of an optical medium, and is more general than the
dynamical Casimir effect, in the sense that it is independent of boundary
conditions, and can occur in unbounded media. On the other hand, the Unruh
effect is quite often related with the dynamical Casimir (see e.g. \cite
{Eberlein1996,Milton2004}), but the nature of this connection has been
debated \cite
{Eberlein1996,Milton2004,Hu2004,Scully2003,Hu-comment,Scully-reply}

One important aspect is that these three mechanisms can create radiation from vacuum with a finite energy, but with zero momentum. Time refraction, in a cavity or in an unbounded medium, creates pairs of photons propagating in opposite directions. The dynamical Casimir effect in a cavity creates a standing wave mode, which is equivalent to two counter-propagating photons. And, it was recently shown that the Unruh emission by accelerated electrons is also made of pairs of photons \cite{habs}, with zero momentum in the instantaneous rest frame. 

Here we discuss quantum vacuum in a variable cavity, and try to establish a
close relation between the three mentioned quantum vacuum processes. A
detailed quantitative and qualitative comparison will be made between these
processes. We hope that such a comparison will help to elucidate the nature
of photon vacuum, and the main properties of a vibrating cavity. In Section
2, we describe the physical configurations of a variable optical cavity and
state the laws of time refraction. In Section 3, we use these laws to
describe the dynamical Casimir effect, and explore further the formal
analogies discussed in our previous work \cite{mendpra}. We show that the
oscillating cavity is unstable to quantum pair creation, and discuss
possible saturation mechanisms for the vacuum instability, including cavity
losses and nonlinear detuning. In Section 4, we introduce the effective
Unruh acceleration, and compare the efficiency of dynamical Casimir in a
vibrating cavity with that of Unruh radiation in unbounded vacuum. Finally,
in Section 5, we state our conclusions.

\section{Time refraction in a cavity}

We first consider an empty cavity with a moving mirror. This is equivalent to an optical cavity with a fixed length, but filled with  a time varying dielectric medium. By changing the refractive index, with the help for instance of an applied external field, we change the optical length of the cavity. Therefore, these two models of a varying cavity, with a variable length or with a varying dielectric medium, are equivalent from the point of view of the optical length.

Let us first relate the temporal change in the refractive index $n (t)$, with the change in the empty cavity length $L (t)$. If we consider a given cavity mode with an integer number $m$ of wavelengths along the cavity axis, corresponding to the wavenumber $k_m = 2 \pi m / L_0$, where $L_0$ is the cavity length, this mode frequency will vary in time according to

\begin{equation}
\omega_m (t) = \frac{k_m c}{n (t)} = \frac{2 \pi m c}{L_0 n (t)} = \frac{2 \pi m c}{L (t)} .
\label{eq:2.1} 
\end{equation}
By writing $n (t) = n_0 + \delta n (t)$, where $n_0$ is the unperturbed refractive index, we obtain for the variable cavity length $L (t) = L_0 + \delta L (t)$, where $\delta L (t) = L_0 \delta n (t)$ is the effective optical displacement.
Keeping this equivalence in mind, we can now focus on the empty cavity case, and adapt previous results obtained for the variable refractive index case. For a single mode in vacuum, we have used an electric field operator of the form

\begin{equation}
\vec{E}_k (x, t) = i \sqrt{\frac{\hbar \omega (t)}{2 \epsilon_0}} \left[ a_k (t) e^{i k x} - a_k^+ (t) e^{- i k x} \right] \vec{e}_k ,
\label{eq:2.2} 
\end{equation}
where $\vec{e}_k$ is the unit polarization vector and the creation and destruction can be written as

\begin{equation}
a_k (t) = A_k (t) e^{- i \int \omega (t) dt} \quad , \quad  a_k^+ (t) = A_k^+ (t) e^{- i \int \omega (t) dt}   .
\label{eq:2.3} \end{equation}
But for a cavity mode $m$, we need to associate the other momentum component $- k$,  and it is more adequate to use the field mode operator 

\begin{equation}
\vec{E}_m (x, t) = i \sqrt{\frac{\hbar \omega_m (t)}{2 \epsilon_0}} \left\{a_m (t) \sin [k_m [t] x] + h.c. \right\} \vec{e}_m ,
\label{eq:2.4} \end{equation}
where $k_m = 2 \pi m / L (t)$. Using a non-perturbative field theory approach, it is then possible to show \cite{mend1,mendpra},  that the mode operators will evolve in time according to the equations
\begin{subequations}
\begin{eqnarray}
&& \frac{d a_m}{d t} = - i \omega_m a_m + \left( \frac{L'}{2 L} \right) a_m^+ ,  
 \\[2mm]&&
\frac{d a_m^+}{d t} =  i \omega_m a_m^+ + \left( \frac{L'}{2 L} \right) a_m ,
\end{eqnarray}
\label{eq:2.5} 
\end{subequations}
where $L' \equiv d L / dt$. The creation and destruction operators can also be represented as

\begin{equation}
a_m (t) = A_m (t) e^{- i \phi (t)} \quad , \quad  a_m^+ (t) = A_m^+ (t) e^{- i \phi (t)}  , 
\label{eq:2.6} \end{equation}
with the phase

\begin{equation}
\phi (t) = \int_0^t \omega_m (t') d t' .
\label{eq:2.6b} \end{equation}
The evolution equations (\ref{eq:2.5}) can then be written in a simpler and more compact form, as

\begin{equation}
\frac{d A_m}{d t} =  \nu (t) A_m^+ 
\quad , \quad 
\frac{d A_m^+}{d t} = \nu (t)^* A_m ,
\label{eq:2.7} \end{equation}
with the coupling function
\begin{equation}
\nu (t) =  \left( \frac{L'}{2 L} \right) \exp [2 i \phi (t)] . 
\label{eq:2.7b} \end{equation}
This system of equations can easily be integrated, leading to the well known solutions
\begin{subequations}
\begin{eqnarray}
&& A_m (t) = \alpha (t) A_m (0) - \beta (t) A_m^+ (0) ,
 \\[2mm] &&
A_m^+ (t) = \alpha (t) A_m^+ (0) - \beta (t) A_m (0) ,
\end{eqnarray}
\label{eq:2.8} 
\end{subequations}
with
\begin{equation}
\alpha (t) = \cosh r (t) \quad , \quad \beta (t) = \sinh r (t)  ,
\label{eq:2.8b} \end{equation}
where the squeezing function $r (t)$ is determined by

\begin{equation}
r (t) = \int_0^t \nu (t') d t' = \frac{1}{2} \int_0^t \left( \frac{d}{d t'} \ln L (t') \right) \exp [2 i \phi (t')] dt'  .
\label{eq:2.9} \end{equation}
Notice that equations (\ref{eq:2.8}) are temporal Bogoliubov relations, obeying the usual hyperbolic condition $\alpha (t)^2 - \beta (t)^2 = 1$, which correspond to bosonic quantum states. These are the quantum laws of time refraction, adapted here to the case of an empty cavity with a variable length $L (t)$. Their physical implications, and their classical counterparts. were discussed in detain in reference \cite{mendpra}.

\section{Dynamical Casimir effect}

Let us now concentrate on the case where we have an oscillating mirror in the empty cavity, as described by

\begin{equation}
L (t) = L_0 + \epsilon \sin (\Omega t) ,
\label{eq:3.1} \end{equation}
where we have assumed that the amplitude of the oscillations is much smaller than the cavity length $\epsilon \ll L_0$. From equation (\ref{eq:2.1}) we can see that the mode frequency will also oscillate in time according to

\begin{equation}
\omega_m (t) = \frac{2 \pi m c}{L (t)} \simeq \omega_{m0} \left[1 - \frac{\epsilon}{L_0} \sin (\Omega t) \right] ,
\label{eq:3.2} \end{equation}
with $\omega_{m0} = k_{m0} c = 2 \pi m c / L_0$. Let us calculate the corresponding values for the functions $\nu (t)$ and $r (t)$. From equation (\ref{eq:2.7b}) we obtain

\begin{equation}
\nu (t) = \frac{\epsilon \Omega}{2 L (t)} \cos (\Omega t) \exp [2 i \omega_{m0} t + i \rho \cos (\Omega t)]  ,
\label{eq:3.3} \end{equation}
with
\begin{equation}
\rho = \frac{2 \epsilon \omega_{m0}}{\Omega L_0}  .
\label{eq:3.3b} \end{equation}
Neglecting higher order corrections with respect to the small parameter $\epsilon / L_0$, we can write

\begin{equation}
\nu (t) = \frac{\rho \Omega^2}{4 \omega_{m0}} \cos (\Omega) e^{2 i \omega_{m0}} \sum_{n = - \infty}^\infty i^n J_n (\rho) e^{i n \Omega t}  ,
\label{eq:3.4} \end{equation}
where $J_n (\rho)$ are Bessel functions of the first kind. Replacing this in the definition of the squeezing function, we can easily see that only the constant terms of $\nu (t)$ will give a significant contribution to $r (t)$, while the others will average out to zero, for $t \gg 1/ \Omega$. Constant terms of $\nu (t)$ only occur for
\begin{equation}
(n \pm 1) \Omega = 2 \omega_{m0}  .
\label{eq:3.4b} \end{equation}
When such a condition is verified, the constant term of $\nu (t)$ is determined by
\begin{equation}
\nu_n = \frac{\rho \Omega^2}{2^3 \omega_{m0}} J_n (\rho)  .
\label{eq:3.5} \end{equation}
For $\rho \sim \epsilon / L_0$, the largest of these terms will correspond to $n = 0$, which leads us to the well known condition for an efficient dynamical Casimir effect, $\Omega = 2 \omega_{m0}$ \cite{dodonov2}. The corresponding expression for the squeezing function is simply determined by

\begin{equation}
r (t) = \nu_0 t = \frac{\epsilon}{L_0} \frac{\omega_{m0}}{2} J_0 (\epsilon/L_0) t  .   
\label{eq:3.6} \end{equation}
Let us now calculate the number of photon pairs created from out of vacuum, due to the oscillations of the cavity mirror. To do this, we start from the usual definition of the photon number operator, for the cavity mode $m$, as given by $N_m (t) = A_m^+ (t) A_m (t)$. The average number of photon pairs created at time $t$ in the cavity will then be determined  by

\begin{equation}
\langle N_m (t) \rangle = \langle 0 | A_m^+ (t) A_m (t) | 0 \rangle ,
\label{eq:3.7} \end{equation}
where $| 0 \rangle$ is the vacuum state vector for the cavity mode $m$. Using the above solutions for the operators (\ref{eq:2.8}), we obtain

\begin{equation}
\langle N_m (t) \rangle = \sinh^2 (\nu_0 t)  .
\label{eq:3.8} \end{equation}
For  short times, such that $\nu_0 t \ll 1$, this equation predicts a linear  growth, given by $\langle N_m (t) \rangle \; \simeq \nu_0 t$.
On the other hand, for very long times, such that $t \gg 1 / \nu_0$, this leads to an exponential growth

\begin{equation}
\langle N_m (t) \rangle \simeq \frac{1}{4} \exp (2 \nu_0 t )  .
\label{eq:3.8b} \end{equation}
This is the well known instability predicted for the dynamical vacuum cavity at the resonant excitation frequency $\Omega = 2 \omega_{m0}$. 
Such an exponential growth is only possible if we neglect the cavity losses, due to absorption and diffraction at the cavity mirrors. These losses can be described by a linear damping rate $\gamma$, such that $Q_{m0} =\omega_{m0} / \gamma$ is the quality factor of the cavity. If we include losses, we can write the following balance equation for the average number of photon pairs

\begin{equation}
\frac{d}{d t} \langle N_m (t) \rangle = 2 \nu_0 \sinh (\nu_0 t) \cosh (\nu_0 t) - \gamma \langle N_m (t) \rangle 
\label{eq:3.9} \end{equation}
For a small damping rate, this is approximately equal to
\begin{equation}
\frac{d}{d t} \langle N_m (t) \rangle = \left[ 2 \nu_0 \coth (\nu_0 t) - \gamma \right] \langle N_m (t) \rangle .
\label{eq:3.10} \end{equation}
This can easily be integrated to give
\begin{equation}
\langle N_m (t) \rangle =  \sinh^2 (\nu_0 t) \exp (- \gamma t) .
\label{eq:3.11} \end{equation}
Let us now consider nonlinear saturation effects, by assuming that the cavity oscillations are produced by an externally driven nonlinear element inserted in the cavity. The existence of photons due to the dynamical Casimir instability will introduce a small change in the refractive index of the dielectric element, as described by the usual law $n (t) = n_0 (t) + n_1 I_m (t)$, where $I_m (t) =  \hbar \omega_{m0} c  \langle N_m (t) \rangle$ is the intensity associated with the cavity mode $m$. The nonlinear refractive index $n_1$ is proportional to the third order susceptibility of the dielectric medium. This nonlinear correction leads to a small detuning of the cavity mode, which can be approximately described by  a nonlinear correction factor of the type  $(1- \zeta \langle N_m (t) \rangle^2)$, with  $\zeta \simeq  (\hbar \omega_{m0} c n_1)^2$, multiplying the driving term of equation (\ref{eq:3.9}). This nonlinear effect will therefore reduce the rate of creation of photons inside the vibrating cavity.

\section{Effective Unruh acceleration}

We now compare the previous quantum effects associated with an oscillating cavity
with those associated with Unruh radiation in unbounded vacuum. According to
Unruh, an observer (or a physical object) moving in vacuum with acceleration 
$a$, will perceive in its accelerated frame, a thermal spectrum with
temperature defined by

\begin{equation}
k_{B}T=\frac{\hbar }{2\pi c}|a|  .
\label{eq:4.1}
\end{equation}
Noting the analogy between moving mirror radiation and the Unruh effect \cite
{obadia}, a single moving cavity wall is treated as
our accelerated observer. Furthermore, this accelerated observer will
interact with a radiation spectrum having the following energy distribution
per field mode \cite{boyer,tajima}
\begin{equation}
W (\omega, t) =  \left[ 1 + \frac{a^2 (t)}{\omega^2 c^2} \right] W_T (\omega, t)  ,
\label{eq:4.2} \end{equation}
where we define the thermal energy distribution
\begin{eqnarray}
&& W_T (\omega, t) = \frac{\hbar}{2} \omega \coth \left( \pi \frac{\omega c}{a (t)} \right) 
\nonumber \\ &&\qquad
= \hbar \omega \left\{ \frac{1}{2} + \frac{1}{\exp [2 \pi \omega c / |a (t)|] - 1} \right\}  .
\label{eq:4.3} 
\end{eqnarray}
This thermal spectrum includes both vacuum fluctuations and Planck spectrum with temperature defined by equation (\ref{eq:4.1}). In the low temperature limit $\hbar \omega \gg k_b T$, or $\omega \gg a (t) c$, it will reduce to the blackbody radiation. 
In alternative, we can use the photons number distribution $N (\omega, t) = W(\omega, t) / \hbar \omega$. Neglecting the unphysical fractional number associated with the vacuum fluctuation term in equation (\ref{eq:4.3}), which represents in fact the average effect of virtual photons, we can write for the Unruh spectrum 

\begin{equation}
N (\omega) =  \left[ 1 + \frac{a^2}{\omega^2 c^2} \right] \frac{1}{\exp [2 \pi \omega c / | a |] - 1}  . 
\label{eq:4.4} \end{equation}
Let us now define  an effective Unruh acceleration $a_{\mathrm{eff}}$, such that, for $\omega = \omega_{m0}$ this nearly thermal spectrum gives the same amount of photons than those produced by the dynamical Casimir effect in a cavity. It will be determined by the equality

\begin{equation}
\langle N_m (t) \rangle  = \left[ 1 + \frac{a_{\mathrm{eff}} (t)^2}{\omega_{m0}^2 c^2} \right] \frac{1}{\exp [2 \pi \omega_{m0} c / | a_{\mathrm{eff}} (t) |] - 1} ,
\label{eq:4.5} 
\end{equation}
where $\langle N_m (t) \rangle$ is determined by equation (\ref{eq:3.8}). This can also be written as

\begin{equation}
\left[ e^{y (t)} - 1 \right] N_c (t) =  1 + \frac{4 \pi^2}{y (t)^2}  ,
\label{eq:4.6} \end{equation}
with
\begin{equation}
N_c (t) =  \frac{\langle N_m (t) \rangle}{V_c} \quad , \quad y (t) = 2 \pi \frac{\omega_{m0} c}{| a_{\mathrm{eff}} (t) |}   .
\label{eq:4.6b} \end{equation}
For moderate values of the effective acceleration, such that $y > 2 \pi$, we simply obtain

\begin{equation}
y (t) = \ln \left[ 1 + \frac{1}{N_c (t)} \right] ,
\label{eq:4.7} \end{equation}
or, in explicit form
\begin{equation}
a_{\mathrm{eff}} (t) = 2 \pi \frac{\omega_{m0} c}{\ln [1 + 1 / N_c (t)]}  .
\label{eq:4.8} \end{equation}
This effective Unruh acceleration associated with the dynamical Casimir effect can then be used to compare the Unruh radiation efficiency for unbounded vacuum with that of cavity vacuum effects. Such a comparison could help in the choice of the more appropriate configurations for quantum vacuum experiments.

In order to establish such a comparison, we can imagine an observer moving in unbounded vacuum with a constant acceleration $a_0 = \epsilon \Omega^2$, equal to the maximum acceleration value attained by the vibrating mirror of a cavity, or
\begin{equation}
a_0 = 4 \epsilon \omega_{m0}^2 = (2 \pi \omega_{m0} c ) \frac{4 \epsilon}{L_0} m  .
\label{eq:4.9} \end{equation}
The corresponding Unruh spectrum observed in unbounded vacuum, would be given by equation (\ref{eq:4.4}) with $a = a_0$. The ratio between the two quantities $a_{\mathrm{eff}} (t)$ and $a_0$, corresponding to effective and  real Unruh acceleration values, is given by
\begin{equation}
R (t) = \frac{a_{\mathrm{eff}} (t)}{a_0} = \frac{L_0}{4 m \epsilon \ln [1 + 1 / N_c (t)]} .
\label{eq:4.10} \end{equation}
The condition $R (t) \gg 1$ can be achieved for large effective values of $a_{\mathrm{eff}}$ such that
\begin{equation}
\ln \left[ 1 + \frac{1}{N_c (t)} \right] \gg \frac{L_0}{4 m \epsilon} .
\label{eq:4.11} \end{equation}
Such a condition can easily be achieved. For instance, assuming that $L_0 \simeq 4 m \epsilon$, which would correspond to $\epsilon \simeq \lambda_m / 4$, where $\lambda$ is the wavelength of the cavity radiation mode $m$, the threshold for high effective acceleration regime, such that $R (t) \geq 1$, is attained for $N_c (t) = (e - 1)^{-1}$. If we take the approximate expression for the number of photons created by the dynamical Casimir effect, equation (\ref{eq:3.8b}), condition (\ref{eq:4.11}) is verified for
\begin{equation}
t \geq \frac{1}{2 \nu_0} \ln \left( \frac{1}{e - 1} \right) \simeq \frac{1}{4 \nu_0} ,
\label{eq:4.12} \end{equation}
This clearly shows that, for $t \gg 1 / \nu_0$, we are in a regime where the dynamical Casimir effect is much more favorable, for quantum vacuum observations, that the corresponding Unruh effect in unbounded vacuum, for an observer moving with the maximum mirror acceleration $a_0 = \epsilon \Omega^2$. This is due to the unstable character of the dynamical Casimir effect.

\section{Summary and conclusions}

In this work we have considered vacuum quantum processes in an oscillating
cavity. Two different quantum processes have been discussed in this specific
configuration: time refraction and dynamical Casimir radiation. We have
shown that time refraction in a cavity is physically identical to the
dynamical Casimir effect, if we identify the varying optical path (due to
the change of the refractive index of the medium inside a cavity with fixed
boundaries) with the actual varying length of an empty cavity with a moving
mirror. We have shown the possible occurrence of an exponential growth of
the number of photons created from vacuum by both models, if the cavity
length oscillates at a frequency $\Omega $ with is twice the frequency of a
given cavity mode $\omega _{m0}$. The instability growth rate is linear with
respect to the mirror displacement $\epsilon $, or to the equivalent
oscillation amplitude of the refractive index $\delta n=\epsilon /L_{0}$.
Previous results obtained for these two models \cite{mendpra} were confirmed
and refined. In particular the instability saturation mechanisms, due to
cavity losses and nonlinear detuning has been included.

We have also introduced the concept of an effective Unruh acceleration for
the dynamical Casimir effect, in order to create a bridge between the
bounded and unbounded quantum vacuum effects. Using this new concept, we
were able to compare the efficiency of the dynamical Casimir effect with
that of an equivalent Unruh radiation, for observers moving in unbounded
vacuum with the maximum value of mirror acceleration $a_{0}=2\epsilon \Omega 
$. We have shown that very efficient regimes for photon creation from vacuum
can be attained in a vibrating cavity, with respect to those of unbounded
vacuum. This suggests that dynamical cavity experiments appear to be better
candidates for the observation of photon creation from a perturbed quantum
vacuum than those in unbounded vacuum, for comparable values of acceleration.

\end{document}